\documentclass[prd,nofootinbib,reprint,twocolumn]{revtex4-1}
\usepackage{amsmath,amssymb,bm,epsfig,color,graphicx}
\usepackage[mathscr]{eucal}
\usepackage{slashed}
\usepackage{cancel}
\usepackage[colorlinks=true,linkcolor=blue,citecolor=blue,urlcolor=blue]{hyperref}
\usepackage[caption=false]{subfig}
\usepackage{hyperref}
\usepackage{tikz,xcolor}
\usepackage{amsmath}
\usepackage{slashed}
\usepackage{comment}
\usepackage{cancel}
\usepackage{multirow}
\usepackage{slashed}

\renewcommand{\thefootnote}{\fnsymbol{footnote}}

\newcommand{\bea}{\begin{array}}
\newcommand{\eea}{\end{array}}
\newcommand{\beq}{\begin{eqnarray}}
\newcommand{\eeq}{\end{eqnarray}}

\newcommand{\MeV}{\  {\rm MeV} }
\newcommand{\GeV}{\  {\rm GeV} }
\newcommand{\TeV}{\  {\rm TeV} }
\newcommand{\1}{\mbox{1}\hspace{-0.25em}\mbox{l}}

\definecolor{orange}{RGB}{255,100,0}
\definecolor{rosepink}{RGB}{248,100,100}

\begin{document}
\rightline{CTPU-PTC-26-24}

\vspace{-0.5cm}

\title{
\vspace{0.5cm}
Dark Transport Erases Baryon Inhomogeneity from Supercooled Phase Transitions
}

\author{
Sudhakantha Girmohanta$^{1}$,\footnote{
E-mail address: sgirmohanta@ibs.re.kr} 
Kohei Kamada$^{2,3,4}$,\footnote{
} 
Yuichiro Nakai$^{5,6}$,\footnote{
E-mail address: ynakai@sjtu.edu.cn}
and Fumio Uchida$^{7}\footnote{
E-mail address: }$\\*[10pt]
$^1${\it \normalsize
Particle Theory and Cosmology Group, Center for Theoretical Physics of the Universe, Institute for Basic Science (IBS), Daejeon, 34126, Korea} \\*[3pt]
$^2${\it \normalsize 
School of Fundamental Physics and Mathematical Sciences, Hangzhou Institute for Advanced Study,
University of Chinese Academy of Sciences (HIAS-UCAS), Hangzhou, 310024, China} \\*[3pt]
$^3${\it \normalsize
International Centre for Theoretical Physics Asia-Pacific (ICTP-AP),
Hangzhou/Beijing, China}  \\*[3pt]
$^4${\it \normalsize 
Research Center for the Early Universe, The University of Tokyo, Bunkyo-ku, Tokyo 113-0033, Japan}  \\*[3pt]
$^5${\it \normalsize 
Tsung-Dao Lee Institute, Shanghai Jiao Tong University, \\ No. 1 Lisuo Road, Pudong New Area, Shanghai 201210, China} \\*[3pt]
$^6${\it \normalsize 
School of Physics and Astronomy, Shanghai Jiao Tong University, \\ 800 Dongchuan Road, Shanghai 200240, China} \\*[3pt]
$^7${\it \normalsize 
Cosmology, Gravity and Astroparticle Physics, Center for Theoretical Physics of the Universe,
Institute for Basic Science (IBS), Daejeon, 34126, Korea
} 
\\*[5pt]
}

\begin{abstract}
Supercooled phase transitions can generate observable gravitational waves, but their bubble dynamics may also imprint large spatial inhomogeneities in the baryon asymmetry. For phase transitions below the TeV scale, ordinary baryon diffusion can be too slow to erase such inhomogeneities before Big Bang Nucleosynthesis, potentially leading to stringent constraints from the observed light-element abundances. We show that this problem can be naturally avoided in darkogenesis scenarios, where the phase transition first generates an asymmetry in a dark-sector particle that is subsequently transferred to visible baryons. The dark-sector particle can travel over much larger distances than ordinary baryons before decaying, erasing the inhomogeneity before the baryon asymmetry is established. We derive the conditions for efficient erasure in both the diffusive and collisionless transport regimes and translate them into constraints on the dark-sector interactions. For a representative dark phase transition with a reheating temperature of $\mathcal{O}(1) \,\mathrm{GeV}$, $\beta/H \simeq 10$, and a dark-sector particle lifetime of $0.1\,\mathrm{s}$, we find that the inhomogeneity is efficiently erased for a viable range of the interactions. Importantly, baryogenesis before the phase transition is challenged both by dilution from strong supercooling and by the subsequent imprinting of spatial inhomogeneities. This points toward darkogenesis associated with the phase transition as a consistent framework, in which dark-sector transport erases the inhomogeneity before it is transferred to visible baryons.
\end{abstract}

\maketitle

\renewcommand{\thefootnote}{\arabic{footnote}}
\setcounter{footnote}{0}

%%%%%%%%%%%%%%%%%%%%%%%%%%%%%%%%%%%%%%%%%%%%%%%%%%%
\section{Introduction}\label{introduction}
%%%%%%%%%%%%%%%%%%%%%%%%%%%%%%%%%%%%%%%%%%%%%%%%%%%

The origin of the baryon asymmetric Universe remains one of the central 
problems in particle physics and cosmology. 
Since any primordial baryon asymmetry is diluted by inflation, the observed  
asymmetry must be generated after inflation and before Big Bang Nucleosynthesis 
(BBN). 
As emphasized by Sakharov, successful baryogenesis requires baryon-number 
violation, C and CP violation, and a departure from thermal equilibrium 
\cite{Sakharov:1967dj}. 
A cosmological first-order phase transition (PT) provides a particularly attractive 
setting for the last requirement, since the nucleation, expansion, and collision 
of bubbles drive the system far from thermal equilibrium.

Strong first-order phase transitions are also of considerable interest as sources of 
stochastic gravitational waves (GWs). 
In particular, the recent evidence for a stochastic GW background at nano-hertz 
frequencies reported by the pulsar timing array (PTA) experiments
\cite{NANOGrav:2023gor,NANOGrav:2023hvm,EPTA:2023fyk,Reardon:2023gzh,Xu:2023wog}
has motivated a cosmological interpretation in terms of a supercooled first-order PT
whose critical temperature is at the QCD-to-GeV scale
\cite{Nakai:2020oit,Bringmann:2023opz,Fujikura:2023lkn,Madge:2023dxc,Megias:2023kiy,Salvio:2023ynn,Salvio:2023blb,Gouttenoire:2023bqy,Addazi:2023jvg,Li:2023bxy,Ghosh:2023aum,Jiang:2023qbm,Wang:2023bbc,Li:2025nja,Fujikura:2025iam}.
Such a PT may naturally occur in a GeV-scale dark sector (DS), which cannot be completely secluded from the Standard Model (SM) sector due to the constraint on the effective number of relativistic species, $\Delta N_{\rm eff}$
\cite{Nakai:2020oit,Bringmann:2023opz}. The required portal interactions connecting the dark and visible sectors then open the possibility of experimentally probing the DS responsible for the GW signal.

Strong supercooling associated with the PT, however, qualitatively changes the cosmological history.
During the supercooling stage, the vacuum energy dominates the energy density of the Universe.
The subsequent release of vacuum energy and reheating can substantially dilute 
any baryon asymmetry and dark matter (DM) abundance produced before the PT.
This observation motivates scenarios in which the observed relic abundances are 
generated during or after the PT
\cite{Fujikura:2024jto,Girmohanta:2025wcq,Girmohanta:2026eqg}.
Darkogenesis
\cite{Shelton:2010ta}
provides an economical realization of this possibility, where an asymmetry is first generated in the DS and is subsequently transferred to visible baryons through a portal interaction, while part of the dark-sector asymmetry remains as asymmetric DM.

Such scenarios, however, face a separate
%There is an additional 
cosmological problem that becomes important for a low-scale supercooled PT 
due to its intrinsically inhomogeneous nature.
It proceeds through the nucleation, expansion and collision of bubbles, and any baryogenesis mechanism tied to this dynamics can therefore generate spatial variations in the baryon asymmetry.
Even when the asymmetry already exists before the PT, the bubble 
dynamics can imprint spatial inhomogeneities on an initially homogeneous 
distribution.
For a sufficiently high-scale transition, subsequent diffusion efficiently 
smooths these fluctuations.
At lower temperatures, however, the diffusion of ordinary baryons becomes too 
slow to erase sufficiently large inhomogeneities before BBN.
Since the primordial light-element abundances depend nonlinearly on the local 
baryon-to-photon ratio, their precise measurements can impose a 
nontrivial constraint on low-scale supercooled phase transitions 
\cite{Applegate:1985qt,Applegate:1987hm,Fuller:1993sp,Heckler:1994uu,Megevand:2004ry,Barrow:2018yyg,Inomata:2018htm,Bagherian:2025puf,Azatov:2026sdm,Bai:2026udq}.
In particular, this issue is relevant to the QCD-to-GeV scale transition
motivated by the PTA signal~\cite{Bringmann:2026xcx, Bagherian:2025puf}.

In this paper, we point out that darkogenesis provides a natural way to evade the baryon-inhomogeneity problem.
The key observation is that the asymmetry remains stored in the DS for a finite period before being transferred to the visible baryons.
During this period, a dark-sector particle $\chi$ carrying the asymmetry can transport it over distances much larger than those accessible to protons and neutrons, provided that its interactions with the thermal plasma are sufficiently weak.
Spatial inhomogeneities generated during the PT can hence be erased before the dark asymmetry is converted into the visible baryon asymmetry.
In this way, the relevant transport scale is set by the dark-sector particle rather than by the much slower diffusion of ordinary baryons.

We explore the dark transport mechanism in a model-independent manner by characterizing the 
interaction of $\chi$ with the thermal plasma through its thermally averaged 
scattering cross section.
We analyze both the diffusive regime, in which the spatial distribution of the dark-sector particle $\chi$ evolves diffusively, and the collisionless regime, in which the erasure is controlled by free streaming. Requiring the characteristic transport length before the decay of $\chi$ to 
exceed the length scale of inhomogeneities generated by the PT 
allows us to derive constraints on the dark-sector interactions.
As a concrete realization, we consider a neutron portal that transfers the 
$\chi$ asymmetry into visible baryons and show the existence of a viable parameter region 
in which baryon inhomogeneities are efficiently erased.

The rest of the paper is organized as follows.
In Sec.~\ref{problem}, we discuss the baryon-inhomogeneity problem associated with 
low-scale supercooled phase transitions.
In Sec.~\ref{resolution}, we investigate the dark transport mechanism and derive the analytic 
conditions under which the inhomogeneity is erased, considering both the 
diffusive and collisionless regimes. We then compare these analytic estimates with a 
detailed numerical evaluation of the relevant length scales, and present a concrete 
example of a dark-sector interaction that satisfies these conditions.
Sec.~\ref{conclusions} is devoted to conclusions and discussions.

%%%%%%%%%%%%%%%%%%%%%%%%%%%%%%%%%%%%%%%%%%%%%%%%%%%
\section{The problem: Inhomogeneity in $\eta$}\label{problem}
%%%%%%%%%%%%%%%%%%%%%%%%%%%%%%%%%%%%%%%%%%%%%%%%%%%
In this section, we quantify the baryon-inhomogeneity problem outlined in 
Sec.~\ref{introduction}. The dynamics of bubble collision imprints inhomogeneity in the baryon density with a correlation length determined by the mean bubble separation, denoted as $R_*$, which is correlated with the Hubble rate at reheating, denoted as $H_{\rm RH}$, namely
\begin{align}
    R_* \sim  (8 \pi)^{1/3} v_w \left(\frac{\beta}{H_{\rm RH}} \right)^{-1} H_{\rm RH}^{-1} \, ,
    \label{Eq:Rst}
\end{align}
where $\beta$ sets the inverse duration of the PT, and $v_w$ represents the bubble wall velocity.\footnote{Note that Eq.~\eqref{Eq:Rst} is usually expressed in terms of the Hubble rate at percolation, denoted as $H_*$. However, for a  supercooled PT, the relevant scale is set by the reheating temperature after the transition, since the vacuum energy dominates during the PT and is converted into radiation upon reheating~\cite{Caprini:2019egz}.} 
The peak frequency of a GW background induced by the PT is correlated with the reheating temperature 
of the PT, namely
\begin{align}
    f_{\rm peak} \simeq 20 {\ \rm nHz} \frac{1}{v_w} \left( \frac{g_*(T_{\rm RH})}{20} \right)^{1/6} \left(\frac{T_{\rm RH}}{\rm GeV} \right) \frac{\beta}{H_{\rm RH}} \, ,
    \label{Eq:GW_peak}
\end{align}
where $f_{\rm peak}$, $g_*$, and $T_{\rm RH}$ denote the peak GW frequency, the effective number of relativistic degrees of freedom and the reheating temperature, respectively.

The created inhomogeneity tends towards homogeneity via a diffusion process, which is inherently slow, given how protons are coupled to the electron-positron plasma via Coulomb interactions, and neutrons via their magnetic dipole moment. Reference \cite{Bagherian:2025puf} argues that since the comoving diffusion lengths of the proton and neutron at the onset of BBN are comparable to the comoving Hubble length at $\simeq 3$ TeV and $\simeq 7$ GeV, respectively, any strongly supercooled PT occurring below these temperatures can leave the protons and/or neutrons inhomogeneous at the onset of BBN.

The resulting baryon inhomogeneity is in conflict with the \%-level precise measurement of the relative deuterium abundance, denoted as D/H, which depends \textit{non-linearly} on the local value of the baryon-to-photon ratio, denoted as $\eta$.\footnote{Depending on the model, the amplitude of the inhomogeneities can be $\delta\eta/\eta=\mathcal{O}(1)$ or even larger if the asymmetry is generated by the bubble dynamics~\cite{Heckler:1994uu,Megevand:2004ry}. Even when a pre-existing baryon asymmetry is only rendered inhomogeneous by the inhomogeneous reheating after the PT, the amplitude can reach $\delta\eta/\eta\sim\mathcal{O}(10)\,H/\beta$~\cite{Bagherian:2025puf}.} This is a consequence of the fact that for the measured value of $\eta$, both the DD and D$p$ channels, namely the $\mathrm{D}+\mathrm{D} \to \mathrm{T}+p$, $\mathrm{D}+\mathrm{D} \to {}^{3}\mathrm{He}+n$, and $\mathrm{D}+p \to {}^{3}\mathrm{He}+\gamma$ have comparable rates, which together determine the final deuterium abundance. Due to this non-linear dependence on $\eta$, the $\%$-level determination of the D/H abundance places constraints on ${\cal{O}}(10\%)$ fluctuations in $\eta$ at the time of BBN~\cite{Inomata:2018htm,Bagherian:2025puf}.\footnote{The precise upper bound on the root-mean-square fluctuation of $\eta$ depends on the adopted nuclear reaction rates~\cite{Mossa:2020gjc}, and is $0.19$ ($0.29$) with \texttt{PArthENoPE} (\texttt{PRIMAT}) according to the latest studies~\cite{Elahi:2026ldn, Launders:2026ciu}.}

The precision measurement of light-element abundances therefore appears to be generically in conflict with a strongly supercooled PT occurring below the TeV scale. In particular, Eq.~\eqref{Eq:GW_peak} implies that a PT with $\beta/H_{\rm RH}\simeq\mathcal{O}(10)$ that generates observable GWs in the nHz-mHz range is constrained by measured light-element abundances. This includes a strong first-order electroweak PT and a PT that may explain the nano-Hz stochastic GW background observed by the PTA collaborations. This raises the important question of how the resulting baryon inhomogeneities can be reconciled with the observed light-element abundances. In the next section, we show that this tension can be resolved by the transport of the asymmetry within the DS before it is transferred to the visible sector.

%%%%%%%%%%%%%%%%%%%%%%%%%%%%%%%%%%%%%%%%%%%%%%%%%%%
\section{Dark transport mechanism}\label{resolution}
%%%%%%%%%%%%%%%%%%%%%%%%%%%%%%%%%%%%%%%%%%%%%%%%%%%

We consider a darkogenesis setup~\cite{Kaplan:2009ag, Shelton:2010ta, Petraki:2013wwa, Zurek:2013wia}, in which a strongly supercooled PT in the DS first creates an asymmetry in a DS Dirac fermion $\chi$ that later decays to visible baryons via a portal, while also transmitting its asymmetry into the DM.
The erasure of inhomogeneity is thus controlled by the transport of $\chi$, which may have a significantly larger diffusion length compared to the visible baryons by virtue of being an SM singlet. As a concrete realization, we consider a dark-sector first-order PT occurring at the QCD-to-GeV scale that explains the PTA signal of GWs while simultaneously providing the observed baryon asymmetry and an asymmetric DM~\cite{Fujikura:2023lkn, Fujikura:2024jto, Girmohanta:2025wcq, Girmohanta:2026eqg}, although our results remain qualitatively unchanged for a broad darkogenesis scenario where the PT occurs below the TeV scale.

We assume that the pre-existing baryon asymmetry and DM are diluted to negligible values after the PT, and that an asymmetry in $\chi$ is generated by the PT dynamics.\footnote{See Refs.~\cite{Fujikura:2024jto, Girmohanta:2025wcq} for explicit examples of such mechanisms.} This asymmetry generically inherits spatial inhomogeneities on the scale of the mean bubble separation $R_*$, with amplitudes that can be $\mathcal{O}(1)$ or larger, as discussed in Sec.~\ref{problem}. If preserved and transferred to the SM baryons, these would be incompatible with inhomogeneous-BBN constraints~\cite{Inomata:2018htm,Bagherian:2025puf}.

In addition, we assume that the reheating temperature of the PT is below the electroweak sphaleron freeze-out temperature ($\simeq 130$ GeV), and that the following neutron portal interaction is responsible for transferring the asymmetry to the visible baryons,
\begin{align}
{\cal O}_{n\chi}
= \frac{1}{\Lambda_n^2}
\left(\overline{\chi^c} d_R^c\right)
\left(\overline{u_R} d_R^c\right) \, ,
\label{Eq:portal}
\end{align}
where $c$ denotes charge conjugation, $u,d$ denote SM up and down quarks, and the color contraction is implicit. To avoid stringent constraints from nucleon decays, the DS fermion $\chi$ is taken to be heavier than the neutron, i.e., $m_\chi \gtrsim 1 \GeV$.\footnote{See Ref.~\cite{Girmohanta:2026eqg} for a UV completion and phenomenological analysis of this operator.} Provided $\Lambda_n \lesssim 100$ TeV, the lifetime of $\chi$, denoted as $\tau_\chi$, is less than $0.1$ s,  which ensures that the neutrino-assisted weak interactions restore the neutron-to-proton ratio to its standard BBN value after $\chi$ decays into visible baryons, allowing BBN to proceed successfully. Furthermore, $\Lambda_n \gtrsim 2 \TeV$ is required to be consistent with collider searches for mono-jet signals, which implies $\tau_\chi \gtrsim 10^{-8} \, {\rm s}$~\cite{ATLAS:2021kxv,Ciscar-Monsalvatje:2023zkk,Girmohanta:2026eqg}. The operator~\eqref{Eq:portal} implies that $\chi$ is an SM singlet and also carries no charge under possible DS gauge symmetries. 

We also consider additional scattering of $\chi$ with the thermal plasma responsible for the diffusion of $\chi$, and parameterize it by a single, model-independent thermally averaged total cross section, denoted as $\langle \sigma v_\mathrm{rel}\rangle_\chi$, where $v_\mathrm{rel}$ denotes the relative velocity between $\chi$ and the scattering partner in the plasma. We will derive the upper bound on this cross section that ensures a sufficiently large diffusion length for $\chi$ to erase the inhomogeneities generated by the PT.
To obtain the most stringent bound on $\langle \sigma v_\mathrm{rel}\rangle_\chi$, we consider $\chi$ scattering with a generic light particle in the thermal bath, denoted as $\varphi$. One can also consider $\chi$ scattering with DM, however, due to the suppressed number density of DM compared to the photon abundance at the relevant timescale, the corresponding constraint is less stringent.

The mean collision time of $\chi$ at temperature $T$ is given by
\begin{align}
{\tau_\mathrm{coll}(T)} = \frac{1}{\mathfrak{n}_\varphi(T) \langle \sigma v_\mathrm{rel}\rangle_\chi} \equiv \frac{M^2}{T^3} \, .
\label{Eq:lmfp}
\end{align}
Here, the physical number density of $\varphi$ is $\mathfrak{n}_\varphi(T)=g_\varphi [\zeta(3)/\pi^2]T^3$, where $g_\varphi$ denotes $\varphi$ internal degrees of freedom including the factor of $3/4$ when $\varphi$ is a fermion, and $T$ denotes the temperature. We assume $\langle \sigma v_\mathrm{rel}\rangle_\chi$ is approximately independent of temperature for the relevant time scale near BBN, which is a valid approximation when $\chi$ becomes non-relativistic. Equation \eqref{Eq:lmfp} thus defines a mass scale denoted $M$, which encodes all model-dependent $\chi$ interactions.

We define the following parameter to distinguish between the diffusive and collisionless regimes of $\chi$ transport,
\begin{align}
\nonumber
 N_{\rm coll} (\tau_\chi)
\equiv \frac{\tau_\chi}{{\tau_\mathrm{coll}}(\tau_\chi)}
& \simeq
\left(\frac{6\times 10^7\,{\rm GeV}}{M}\right)^2 \\
& \times \left(\frac{0.1\,{\rm s}}{\tau_\chi}\right)^{1/2}
\left(\frac{10}{g_*(\tau_\chi)}\right)^{3/4} \, , 
\label{Eq:Ncoll}
\end{align}
which is the number of collisions per Hubble time, evaluated at $\chi$ decay.
Note that a single collision with $\varphi$ does not affect the ballistic motion of a $\chi$ particle, since each momentum transfer is of the order of $T$, which is much smaller than the momentum that a $\chi$ particle has $\sim \sqrt{m_\chi T}$. We need $m_\chi /T$ 
collisions to change the direction, taking into account the random walk in changing their direction.

When $N_{\rm coll}(\tau_\chi) \gtrsim \max[m_\chi/T(\tau_\chi),1]$, where $T(\tau_\chi)$ denotes the temperature at time $\tau_\chi$, $\chi$ undergoes a sufficient number of collisions with relativistic $\varphi$ particles in the thermal plasma for the diffusion treatment to be valid. Since $N_{\rm coll}(t)\propto t^{-1/2}$ decreases while the threshold $m_\chi/T(t)\propto t^{1/2}$ increases with time, $\chi$ then remains in the diffusive regime throughout its lifetime. On the other hand, for $N_{\rm coll}(\tau_\chi)<1$, $\chi$ is in the collisionless regime at the time of its decay, which we treat separately. The actual boundary between the two regimes lies somewhere in the range $1 \lesssim N_\mathrm{coll}(\tau_\chi) \lesssim m_\chi/T(\tau_\chi)$. For example, for $m_\chi \sim 1$ GeV and $\tau_\chi\simeq 0.1$ s, this corresponds to $3 \times 10^6 \GeV \lesssim M\lesssim 6\times10^7 \GeV$, or equivalently $10^{-6}\,{\rm pb} \lesssim \langle\sigma v_\mathrm{rel}\rangle_\chi\lesssim 4 \times 10^{-4}\,{\rm pb}$ for $g_\varphi=1$.

We also define the total number of collisions experienced by $\chi$ between reheating and its decay as
\begin{align}
    N_{\rm tot} \equiv \int_{t_{\rm RH}}^{\tau_\chi} \frac{dt}{\tau_\mathrm{coll}(t)} & \simeq 2\, \frac{T_\mathrm{RH}-T(\tau_\chi)}{T(\tau_\chi)}\, \frac{\tau_\chi}{\tau_\mathrm{coll}(\tau_\chi)} \notag \\
    & \simeq 2\left[N_\mathrm{coll}(t_\mathrm{RH}) - N_\mathrm{coll}(\tau_\chi)\right] ,
    \label{Eq:Ntot}
\end{align}
where we have neglected the variation of $g_*$ with time. Since $N_{\rm coll}(t_{\rm RH})\gg N_{\rm coll}(\tau_\chi)$, the collisions occur mainly around reheating. Therefore, for $N_{\rm tot} \gtrsim \max[m_\chi/T_{\rm RH},1]$, $\chi$ undergoes enough collisions around reheating to acquire a thermal velocity dispersion through its interaction with $\varphi$ alone.

\subsection{Estimation of the Erasure Length in the Diffusive Regime}
\label{section:diffusiveTreatment}

The erasure of the inhomogeneity of $\chi$ can be described by
the Boltzmann equation for the phase-space distribution of $\chi$, denoted by $f(\vec{x}_c,\vec{p}_c,t)$, which can be written as
\begin{align}
\frac{\partial f}{\partial t}
+ \frac{v(p_c,t)}{a(t)}\,\hat{p}_c \cdot\nabla_{{x}_c} f
= C[f]
-\frac{1}{\tau_\chi}\frac{m_\chi}{E_\chi({p}_c,t)}f \, ,
\label{Eq:phaseSpace}
\end{align}
where $\vec{x}_c$ denotes the comoving coordinate, $a(t)$ is the scale factor, $\vec{p}_c$ is the comoving momentum, which is related to the physical momentum $\vec{p}$ through $p_c\equiv |\vec{p}_c|=a|\vec{p}| = ap$, and $\hat{p}_c \equiv \vec{p}_c/p_c$. Here, $C[f]$ encodes all collision terms for $\chi$, while the last term represents the decay of $\chi$, including the appropriate boost factor determined by the physical energy $E_\chi(|\vec{p}_c|,t)\equiv(p_c^2/a(t)^2+m_\chi^2)^{1/2}$, while $v\equiv p_c/(a(t)E_\chi)$ denotes the peculiar velocity. 

When $N_{\rm coll}(t)\gtrsim\max[m_\chi/T(t),1]$, $f$ remains in local kinetic equilibrium, and integrating over the comoving momentum yields the relevant Boltzmann equation for the comoving number density $n_\chi(\vec{x}_c,t)\equiv\int d^3 \vec{p}_c f(\vec x_c,\vec p_c,t)$. We decompose the comoving number density into a homogeneous and an inhomogeneous component as
\begin{align}
n_\chi(\vec{x}_c,t)
= \bar{n}_\chi(t)
+ \delta n_\chi(\vec{x}_c,t) \ .
\end{align}
Defining the fractional perturbation as
\begin{align}
\varepsilon_\chi(\vec{x}_c,t)
= \frac{\delta n_\chi(\vec{x}_c,t)}{\bar{n}_\chi(t)} \, ,
\label{Eq:epsilonDef}
\end{align}
one obtains
\begin{align}
\partial_t\varepsilon_\chi(\vec{x}_c,t)
= \frac{D_\chi(t)}{a(t)^2}
\nabla_{{x}_c}^2\varepsilon_\chi(\vec{x}_c,t) \, ,
\label{Eq:epsilonChi}
\end{align}
where $D_\chi(t)$ denotes the diffusion coefficient of $\chi$, which is determined by $C[f]$ and hence by $\langle \sigma v_\mathrm{rel} \rangle_\chi$, or equivalently $M$. Since $\chi$ is heavier than a neutron, it is a good approximation to take it to be non-relativistic for the relevant epoch. In this regime, the diffusion coefficient is estimated by
\begin{align}
   D_\chi(t)\sim \langle v_\chi^2 \rangle \, \left[ \frac{m_\chi}{T} \tau_\mathrm{coll}(t) \right]  \sim \tau_\mathrm{coll}(t) =\frac{M^2}{T(t)^3} \, , 
   \label{Eq:Dchitcoll}
\end{align}
where $\langle v_\chi^2\rangle \sim T/m_\chi$ is the thermal velocity dispersion of $\chi$, and the bracket encodes the fact that $m_\chi/T$ collisions are needed to change $\chi$ momentum by order one. 

In Eq.~\eqref{Eq:epsilonChi}, $N_{\rm coll}(t)\gtrsim\max[m_\chi/T(t),1]$ ensures the validity of Fick's relation~\cite{Lifshitz1981Physical}, which relates the comoving flux $\mathbf{J}$ to the density gradient via
\begin{align}
    \nonumber
    \mathbf{J}(\vec{x}_c,t) &\equiv \frac{1}{a(t)} \int d^3\vec{p}_c \, v(p_c,t) \, \hat{p}_c \, f(\vec{x}_c,\vec{p}_c,t) \\
    &= -\dfrac{D_\chi(t)}{a(t)^2}\, \nabla_{x_c} n_\chi(\vec{x}_c,t) \, .
    \label{Eq:FicksLaw}
\end{align}
This relation no longer holds in the collisionless regime, which we treat separately in the next subsection. 

Transforming to Fourier space, and expressing in terms of the physical wavenumber ${k}={k}_c/a$, the solution to Eq.~\eqref{Eq:epsilonChi} is
\begin{align}
\nonumber
\varepsilon_\chi(\vec{k},t)
&= \varepsilon_\chi(\vec{k},t_{\rm RH})
\exp\left[
-{k}^2 \, a(t)^2
\int_{t_{\rm RH}}^t
\frac{D_\chi(t')}{a(t')^2}\,dt'
\right] \\
& = \varepsilon_\chi(\vec{k},t_{\rm RH}) \exp[-{k}^2 \, L_{\rm erase}^2(t)] \, .
\label{Eq:epsilonSol}
\end{align}
It is clear from Eq.~\eqref{Eq:epsilonSol} that any inhomogeneity encoded in the initial perturbation $\varepsilon_\chi(\vec{k},t_{\rm RH})$ is exponentially erased on scales below the physical erasure length $L_{\rm erase}$, defined as
\begin{align}
L_{\rm erase}^2(t)
= a(t)^2
\int_{t_{\rm RH}}^t
\frac{D_\chi(t')}{a(t')^2}\,dt' \ .
\label{Eq:Lerase}
\end{align}
Therefore, the initial inhomogeneity imprinted on length scale $R_*$ at $t_{\rm RH}$ will be erased by $\chi$ diffusion provided
\begin{align}
    L_{\rm erase}(\tau_\chi) \gtrsim \frac{a(\tau_\chi)}{a(t_{\rm RH})} R_* \, .
    \label{Eq:eraseCond}
\end{align}

From Eq.~\eqref{Eq:lmfp}, it is evident that the integrand in Eq.~\eqref{Eq:Lerase} scales as $1/T$, and is therefore dominated by diffusion at late times. An analytical estimate of the erasure length by the time $\chi$ decays into baryons can thus be obtained as
\begin{align}
L_{\rm erase}(\tau_\chi)
\simeq \sqrt{\tau_\mathrm{coll}(\tau_\chi)\,\tau_\chi} = \frac{M}{T(\tau_\chi)} \sqrt{\frac{\tau_\chi}{T(\tau_\chi)}} \, . 
\label{Eq:LeraseAna}
\end{align} 
Assuming radiation domination, we use the time-temperature relation
\begin{align}
    T(t) \simeq \left( \frac{45}{2 \pi^2 g_*(T)}\right)^{1/4} \sqrt{\frac{M_{\rm pl}}{t}} \, ,
    \label{Eq:timeTemp}
\end{align}
where $M_{\rm pl} = 2.4 \times 10^{18}$ GeV is the reduced Planck mass. 

Utilizing Eqs.~\eqref{Eq:Rst},~\eqref{Eq:lmfp},~\eqref{Eq:LeraseAna}, and~\eqref{Eq:timeTemp}, the erasure condition in Eq.~\eqref{Eq:eraseCond}, required for consistency with BBN, can be recast as a lower bound on the scale $M$, namely
\begin{align}
    \nonumber
    M & \gtrsim 65 \ {\rm TeV} \left( \frac{0.1\,{\rm s}}{\tau_\chi} \right)^{3/4} \left( \frac{\beta/H_{\rm RH}}{10}\right)^{-1} \left( \frac{1 {\ \rm GeV}}{T_{\rm RH}}\right) \\
   & \times \left(\frac{20}{{g_*(T_{\rm RH})}}\right)^{1/2} \left(\frac{10}{g_{*}(\tau_{\chi})}\right)^{1/8} \left( \frac{g_{*s}(T_{\rm RH})}{g_{*s}(\tau_\chi)} \right)^{1/3} \, ,
   \label{Eq:Mbound}
\end{align}
where $g_{*s}$ denotes the degrees of freedom for the entropy density. $M \gtrsim 65$ TeV translates to $\langle \sigma v_\mathrm{rel} \rangle_\chi \lesssim 0.8 \,$pb for $g_\varphi=1$. For concreteness, we fix $v_w\simeq 1$ throughout, as appropriate for a strongly supercooled PT. 

%%%%%%%%%%%%%%%%%%%%%%%%%%%%%%%%
\begin{figure}[t!]
    \centering
    \includegraphics[width=0.99\linewidth]{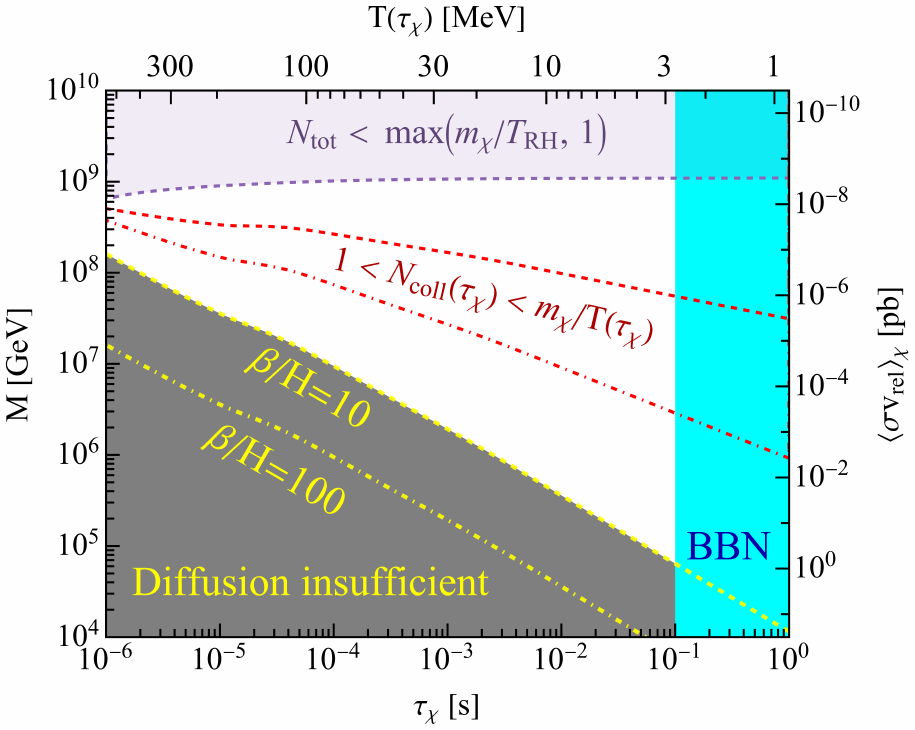}
    \caption{Constraint on the scale $M$, defined through the total scattering cross-section for $\chi$ in Eq.~\eqref{Eq:lmfp}, as a function of the $\chi$ lifetime $\tau_\chi$, assuming $T_{\rm RH}=1$ GeV. The right-hand axis gives the corresponding $\langle\sigma v_\mathrm{rel}\rangle_\chi$ in pb for $g_\varphi=1$, and the top axis the photon temperature $T(\tau_\chi)$, obtained from Eq.~\eqref{Eq:timeTemp} with $g_*(T)$, $g_{*s}(T)$ from Ref.~\cite{Saikawa:2018rcs}. In the gray region, the erasure condition Eq.~\eqref{Eq:eraseCond} is not satisfied for $\beta/H_{\rm RH} = 10$, while a higher $\beta/H_{\rm RH}$ (or $T_{\rm RH}$) relaxes the bound, as shown by the yellow dot-dashed contour for $\beta/H_{\rm RH}=100$. The cyan region is excluded by BBN due to late injection of baryon number~\cite{Girmohanta:2026eqg}. Between the red dashed and dot-dashed lines, $1<N_{\rm coll}(\tau_\chi)<m_\chi/T(\tau_\chi)$ for $m_\chi=1\GeV$, indicating a smooth cross-over, with the diffusive treatment valid below and the collisionless one above. Below the purple region, $\chi$ undergoes enough collisions shortly after reheating [Eq.~\eqref{Eq:Ntot}] to acquire a thermal velocity dispersion, which is sufficient for erasure in the free-streaming regime (Sec.~\ref{sec:collisionless}). In the purple region, the erasure becomes model-dependent, depending on how the $\chi$ momentum distribution is set up, e.g.\ during bubble collisions (Sec.~\ref{sec:baryondamp}). In the white region, $\chi$ transport alone suffices to erase the inhomogeneity before $\chi$ decays into visible baryons.}
    \label{fig:Mbnd}
\end{figure}
%%%%%%%%%%%%%%%%%%%%%%%%%%%%%%%%%%%%%%%%%%%%

The erasure condition in Eq.~\eqref{Eq:eraseCond} is depicted in Fig.~\ref{fig:Mbnd} for a benchmark value of $T_{\rm RH}=1$ GeV, where the gray region is excluded and its boundary is shown by the yellow dashed (dot-dashed) line for $\beta/H_{\rm RH}=10$ ($100$). We use the numerical data for the effective degrees of freedom as a function of temperature from Ref.~\cite{Saikawa:2018rcs}. The boundary follows the approximate $\tau_\chi^{-3/4}$ scaling of Eq.~\eqref{Eq:Mbound}, namely, for larger $\tau_\chi$, $\chi$ has more time to diffuse before decaying, which relaxes the bound. The non-trivial structure of the contours at $\tau_\chi \lesssim 10^{-4} \, {\rm s}$ arises from the rapid variation of the effective degrees of freedom around the QCD scale, which corresponds to the relevant timescale set by $\tau_\chi$. The bound also becomes weaker as $\beta/H_{\rm RH}$ increases. This can be understood from the fact that a larger $\beta/H_{\rm RH}$ implies a faster completion of the PT and, consequently, a smaller mean bubble separation. The resulting inhomogeneities are therefore generated on smaller length scales, making them easier to erase through diffusion. Similarly, increasing $T_{\rm RH}$ weakens the bound, since the Hubble radius at reheating decreases, leading to a smaller characteristic scale for the initial inhomogeneity. In the purple region, $N_{\rm tot}\lesssim\max[m_\chi/T_{\rm RH},1]$, so scattering with $\varphi$ alone does not establish a thermal velocity dispersion after reheating. Erasure there is not excluded, but becomes model-dependent. As discussed in Sec.~\ref{sec:baryondamp}, a power-law suppression of the baryon inhomogeneity persists as long as the $\chi$ momenta are isotropic and not concentrated near $p_c\simeq0$.

%%%%%%%%%%%%%%%%%%%%%%%%%%%%%%%%%%%%%%%%%%%%%%%%%%%%%%%
\subsection{Estimation of the Erasure Length in the Free-streaming Regime}
%for $N_{\rm coll} < 1$}
\label{sec:collisionless}
%%%%%%%%%%%%%%%%%%%%%%%%%%%%%%%%%%%%%%%%%%%%%%%%%%%%%%%

We now consider the regime when $N_{\rm coll} (\tau_\chi) < 1$. In this case, one can ignore the collision term in Eq.~\eqref{Eq:phaseSpace} after $\chi$ decouples, and the Boltzmann equation reduces to
\begin{align}
    \frac{\partial f}{\partial t}
+ \frac{v({p}_c,t)}{a(t)}\,\hat{p}_c \cdot\nabla_{{x}_c} f
= 
-\frac{1}{\tau_\chi}\frac{m_\chi}{E_\chi({p}_c,t)}f \, .
\label{Eq:collisionLess}
\end{align}
Furthermore, the phenomenological relation in Eq.~\eqref{Eq:FicksLaw} is no longer valid for $N_{\rm coll}<1$, and Eq.~\eqref{Eq:collisionLess} has to be solved for $f$ itself. Eq.~\eqref{Eq:collisionLess} implies that $f$ is carried along the following trajectory
\begin{align}
    \nonumber
    f(\vec{x}_c,\vec{p}_c,t) & =f(\vec{x}_c-\hat{p}_c \, \ell_{\rm fs}(p_c,t),\vec{p}_c,t_{\rm RH}) \\
    & \times \exp\left[-\int_{t_{\rm RH}}^t \frac{dt'}{\tau_\chi} \frac{m_\chi}{E_\chi({p}_c,t')}\right] \ ,
    \label{Eq:fSolCollLess}
\end{align}
where the comoving free-streaming length $\ell_{\rm fs}$ is defined as
\begin{align}
    {\ell_{\rm fs}}(p_c,t) = \int_{t_{\rm RH}}^{t} \frac{v(p_c,t')}{a(t')} dt' \, .
    \label{Eq:lfs}
\end{align}

For simplicity, we now neglect the momentum dependence of $E_\chi$ in Eq.~\eqref{Eq:fSolCollLess}, since $\chi$ is non-relativistic. In this case, given an initial inhomogeneity described by $f(x_c,p_c,t_{\rm RH}) = f_0(p_c) \left[ 1+\varepsilon_\chi(x_c,t_{\rm RH}) \right]$, the advection in Eq.~\eqref{Eq:fSolCollLess} becomes a phase in Fourier space, and one obtains the following solution for the fractional perturbation, as defined in Eq.~\eqref{Eq:epsilonDef}, namely
\begin{align}
    \nonumber
    \frac{\varepsilon_\chi(k_c,t)}{\varepsilon_\chi(k_c,t_{\rm RH})} &= \frac{\int d^3 \vec{p}_c \, f_0(p_c) e^{-ik_c\cdot\hat{p}_c \ell_{\rm fs}(p_c,t)}}{\int d^3 \vec{p}_c \, f_0(p_c)} \\
    &=\frac{\int d p_c \, p_c^2 \, f_0(p_c) j_0(k_c \, \ell_{\rm fs}(p_c,t) )}{\int d {p}_c \, p_c^2 \, f_0(p_c)} \ ,
    \label{Eq:BesselSol}
\end{align}
where the angular integration of $e^{-ik_c\cdot\hat{p}_c \ell_{\rm fs}(p_c,t)}$ results in the zeroth-order spherical Bessel function of the first kind, namely $j_0(x)=\sin(x)/x$, and the exponential decay factor in Eq.~\eqref{Eq:fSolCollLess} drops out of the ratio.

Let us now specialize to the case where $f_0(p_c)$ is given by a Maxwell-Boltzmann (MB) distribution, namely
\begin{align}
    f_0(p_c) = {\cal N} \exp \left(-\frac{p_c^2}{2 \sigma_{p_{c}}^2} \right) \ ,
    \label{Eq:MBDist}
\end{align}
with normalization ${\cal N}$, and the variance in the comoving momentum $\sigma_{p_c} = a(t_{\rm RH}) \sqrt{m_\chi T_{\rm RH}}$, 
which is expected in the case where $\chi$ particles were once in kinetic equilibrium just after reheating. In the non-relativistic case, the comoving free-streaming length in Eq.~\eqref{Eq:lfs} becomes
\begin{align}
    {\ell}_{\rm fs}(p_c,t) = \frac{p_c}{m_\chi} \int_{t_{\rm RH}}^{t} \frac{dt'}{a^2(t')} \ .
    \label{Eq:NonRelLFS}
\end{align}

Substituting Eqs.~\eqref{Eq:MBDist}, \eqref{Eq:NonRelLFS} into Eq.~\eqref{Eq:BesselSol}, and switching to the physical wavenumber $k=k_c/a$, we again obtain an exponential form
\begin{align}
    {\varepsilon_\chi(k,t)} &= {\varepsilon_\chi(k,t_{\rm RH})} \exp\left[ - k^2 \, d_{\rm fs}^2(t) \right] \, ,
    \label{Eq:epsilonSol-fs}
\end{align}
where 
\begin{align}
    \nonumber
    d_{\rm fs}(t) &= a(t)\, a(t_{\rm RH}) \sqrt{\frac{T_{\rm RH}}{2 m_\chi}} \int_{t_{\rm RH}}^{t} \frac{dt'}{a(t')^2} \\
    & \simeq \sqrt{\frac{T_{\rm RH}}{2 m_\chi}} \sqrt{t_{\rm RH} \, t} \, \ln\left( \frac{ t}{t_{\rm RH}} \right) \, \ , \label{Eq:FreeStreamingLength}
\end{align}
and we have ignored the subleading term due to variation of $g_{*s}$. 

As in the diffusive case, since the inhomogeneity of $\chi$ is exponentially suppressed on scales smaller than $d_\mathrm{fs}$, the condition for erasing the inhomogeneity before the onset of BBN reads
\begin{align}
    d_{\rm fs}(\tau_\chi) \gtrsim \frac{a(\tau_\chi)}{a(t_{\rm RH})} R_* \, ,
    \label{Eq:ErasureCondCollLess}
\end{align}
where $d_{\rm fs}$ is evaluated at $\tau_\chi$, when the suppression factor in Eq.~\eqref{Eq:fSolCollLess} becomes significant and $\chi$ transfers its asymmetry to the visible baryons.
The condition in Eq.~\eqref{Eq:ErasureCondCollLess} is satisfied when
\begin{align}
    \left( \frac{\beta/H_{\rm RH}}{10} \right) \sqrt{\frac{T_{\rm RH}}{m_\chi}} \ln\left( \frac{\tau_\chi}{t_{\rm RH}} \right) \gtrsim 1 \, ,
    \label{Eq:AnaCollLess}
\end{align}
which is easily satisfied as long as $\tau_\chi \gg t_{\rm RH}$. Numerically, for $\beta/H_{\rm RH} =10$ and $T_{\rm RH}\simeq m_\chi \simeq 1$ GeV, it requires $\tau_\chi \gtrsim 10^{-6}\, {\rm s}$. Therefore, the inhomogeneity is erased in the white region above the red dashed line in Fig.~\ref{fig:Mbnd}.\footnote{We have assumed that $\chi$ enters the free-streaming regime shortly after reheating, with its velocity dispersion set at $T_{\rm RH}$. If instead $\chi$ is initially in the diffusive regime and only later enters the free-streaming regime, $T_\mathrm{RH}$ and $t_\mathrm{RH}$ in Eqs.~\eqref{Eq:FreeStreamingLength} and~\eqref{Eq:AnaCollLess} should be replaced by the temperature and time of this transition. Our conclusion nevertheless remains unchanged, since the diffusion prior to the transition further contributes to the erasure.}

We note that it is the dispersion in the $\chi$ velocity vectors, rather than the velocity itself, that sets the erasure scale, namely, if all $\chi$ particles shared a common velocity, the inhomogeneity would simply be translated rigidly, without any damping. This is reflected in Eq.~\eqref{Eq:BesselSol}. Below the purple region in Fig.~\ref{fig:Mbnd}, $N_{\rm tot}\gtrsim\max[m_\chi/T_{\rm RH},1]$, so that $\chi$ scatters sufficiently around reheating to acquire a thermal velocity dispersion and enters the collisionless regime soon afterwards. For a GeV-scale PT this requires $\langle\sigma v_{\rm rel}\rangle_\chi\gtrsim\mathcal{O}(10^{-9})\,{\rm pb}$. This should be understood as a sufficient rather than a necessary condition. Without a thermal dispersion, the exponential damping of the $\chi$ distribution is lost, but a power-law suppression of the baryon inhomogeneity can still persist, as discussed in Sec.~\ref{sec:baryondamp}. 
In the purple region, the erasure therefore depends on how the $\chi$ momentum distribution is set up, and is model-dependent.

%%%%%%%%%%%%%%%%%%%%%%%%%%%%%%%%%%%%%%%%%%%%%%%%%%%%%%%
\subsection{Beyond the Sudden-Decay Approximation: Comparison with Analytical Estimates}
\label{sec:baryondamp}
%%%%%%%%%%%%%%%%%%%%%%%%%%%%%%%%%%%%%%%%%%%%%%%%%%%%%%%

In the previous subsections, we have shown that the relative inhomogeneity $\varepsilon_\chi(k,t)$ of $\chi$ particles is damped at small scales in both the diffusive [Eq.~\eqref{Eq:epsilonSol}] and the free-streaming [Eq.~\eqref{Eq:epsilonSol-fs}] regimes. Accordingly, we have analytically obtained the conditions on the model-dependent parameters, Eqs.~\eqref{Eq:Mbound} and \eqref{Eq:AnaCollLess}, by comparing the inhomogeneity scale $R_*$ with the damping scale of $\varepsilon_\chi(k,t)$ evaluated at $\tau_\chi$. This comparison implicitly assumes the sudden-decay approximation, namely that $\chi$ decays sharply at $t=\tau_\chi$, so that the baryon distribution at BBN simply inherits the inhomogeneity of $\chi$ at that moment. In practice, however, $\chi$ decays continuously from $t_{\rm RH}$ onward. Baryons produced at earlier times inherit the $\chi$ inhomogeneity before it has been fully damped, and subsequently evolve only through the much slower baryon diffusion. In this subsection, we quantify this effect by numerically solving for the baryon distribution sourced by the continuous decay of $\chi$, and compare the resulting damping scale with our analytic estimates.

After $\chi$ decays, neutron diffusion introduces an additional damping of the baryon distribution, so that the suppression factor becomes $\exp[-k_c^2(L_c(\tau_\chi)^2+d_n(T)^2/2)]$ for $T\lesssim T(\tau_\chi)$. Here the comoving erasure length is defined as $L_c(t)\equiv L_{\rm erase}(t)/a(t)$ in the diffusive regime and $L_c(t)\equiv d_{\rm fs}(t)/a(t)$ in the free-streaming regime, while $d_n(T)$ denotes the comoving neutron damping scale at temperature $T$. This additional damping is negligible, since $d_n\ll L_c(\tau_\chi)$ throughout the parameter region satisfying the bound in Fig.~\ref{fig:Mbnd}.

We now evaluate the damping scale of the baryon inhomogeneity, taking into account both the continuous decay of $\chi$ into neutrons and the subsequent diffusion of the SM baryons. As shown in Secs.~\ref{section:diffusiveTreatment} and~\ref{sec:collisionless}, the comoving number density of $\chi$ in Fourier space evolves as
\begin{align}
    n_\chi(\vec k_c, t)
    =&\,n_\chi(\vec k_c, t_{\rm RH})\notag\\
    &\times\exp\left(
        -k_c^2L_c(t)^2-\frac{t-t_{\rm RH}}{\tau_\chi}
    \right) ,
    \label{eq:nchi}
\end{align}
where the factor $\exp[-(t-t_{\rm RH})/\tau_\chi]$ describes the decay of non-relativistic $\chi$, and hence tracks the baryons continuously produced since $t_{\rm RH}$. This factor cancels between the numerator and denominator of the fractional perturbation $\varepsilon_\chi$ in Eqs.~\eqref{Eq:epsilonSol} and~\eqref{Eq:epsilonSol-fs}, but must be retained here. The decay of $\chi$ then acts as a source term in the Boltzmann equations for the comoving neutron and proton number densities
\begin{align}
    \partial_t\left(\begin{matrix}n_n\\n_p\end{matrix}\right)
    = \hat{D}_{\rm SM}\left(\begin{matrix}n_n\\n_p\end{matrix}\right)+\frac{n_\chi(\vec k_c,t)}{\tau_\chi}\left(\begin{matrix}{\rm Br}\\1-{\rm Br}\end{matrix}\right) \, ,
    \label{eq:dif-evo-bar}
\end{align}
where $n_{n/p}(\vec k_c,t)$ are the comoving neutron and proton number densities in Fourier space, and ${\rm Br} \equiv {\cal B}(\chi \to n+\cdots)$ is the branching ratio of $\chi$ decays into neutrons. We take ${\rm Br} \simeq 1$ for simplicity, which holds for $m_n < m_\chi < m_p+m_{\pi^-}$, where $\chi \to p e^{-} \bar{\nu}_e$ is suppressed by the weak interaction, while the analysis extends straightforwardly to general ${\rm Br}$ following Ref.~\cite{Girmohanta:2026eqg}. The matrix $\hat{D}_{\rm SM}$ collectively describes the relevant SM processes, namely the weak interactions and the diffusion due to neutron-electron, neutron-proton, and proton-electron scatterings~\cite{Applegate:1987hm,Bagherian:2025puf}. Assuming instantaneous decoupling of the weak interactions at $T_{\rm dec}\simeq 0.8\,{\rm MeV}$, Eq.~\eqref{eq:dif-evo-bar} reduces to~\cite{Applegate:1987hm}
\begin{align}
    &\dfrac{n_n}{n_n+n_p}=X_n(t)\,,\\
    &\partial_t n_n=\tilde X_n(t)\left[-k_c^2a(t)^{-2}D_n(t) n_n+\dfrac{n_\chi(\vec k_c,t)}{\tau_\chi}\right] \, ,
    \label{eq:dif-evo-neu}
\end{align}
where $X_n$ is the neutron fraction among SM baryons, and $\tilde X_n$ is the fraction of time a baryon spends as a neutron. While the weak interactions are in equilibrium, $T(t)\geq T_{\rm dec}$, rapid $n\leftrightarrow p$ interconversion implies that each baryon spends a fraction $\tilde X_n(t)=X_n(t)=1/(1+e^{Q/T(t)})$ of its time as a neutron, with $Q=m_n-m_p=1.29\,{\rm MeV}$. Since only neutrons diffuse efficiently, both the source and the diffusion terms are weighted by this factor. After the weak interactions freeze out, $T(t)<T_{\rm dec}$, a neutron remains a neutron, so that $\tilde X_n(t)=1$, while $X_n(t)\simeq1/6$. The neutron diffusion coefficient $D_n(t)$ parametrizes a subset of the SM interactions as $\hat D_{\rm SM}(k_c,t)_{ij}\supset [-k_c^2a(t)^{-2}D_n(t)]\delta_{in}\delta_{jn}$, where the indices run over $n$ and $p$. $D_n(t)$ is dominated by neutron-electron and neutron-proton scatterings, and we adopt the formula for it from Ref.~\cite{Bagherian:2025puf}.

Substituting Eq.~\eqref{eq:nchi} into Eq.~\eqref{eq:dif-evo-neu}, and assuming no initial neutron inhomogeneity, we obtain
\begin{align}
    n_n(\vec k_c,t)={\cal T}_{n\chi}(k_c,t,t_{\rm RH}) \, n_\chi(\vec k_c,t_{\rm RH}) \, ,
    \label{eq:nn-sol}
\end{align}
where we have defined the transfer function
\begin{align}
    &\!\!{\cal T}_{n\chi}(k_c,t,t_{\rm RH})\notag\\
    &\!\,\equiv\int_{t_{\rm RH}}^t\!\!dt'\,\dfrac{\tilde X_n(t')}{\tau_\chi}
    \exp\left[-k_c^2\tilde L_c(t,t')^2-\frac{t'-t_{\rm RH}}{\tau_\chi} \right] \, ,
    \label{eq:def-trf}\\
    &\!\!\tilde L_c(t,t')^2\equiv L_c(t')^2+\int_{t'}^tdt''\,\tilde X_n(t'')\,a(t'')^{-2} D_n(t'') \, . 
    \label{eq:ltilde}
\end{align}
These equations admit a simple interpretation. At each time $t'$, the surviving $\chi$ population is reduced by the factor $e^{-(t'-t_{\rm RH})/\tau_\chi}$, and its inhomogeneity has already been damped on the scale $L_c(t')$. The decays at $t'$ then source baryons, weighted by the fraction of time $\tilde X_n$ they spend as neutrons. These neutrons continue to diffuse until the evaluation time $t$, as encoded in the second term of $\tilde L_c(t,t')$. This contribution is subdominant whenever neutron diffusion is much less efficient than that of $\chi$, in which case $\tilde L_c(t,t')\simeq L_c(t')$.

%%%%%%%%%%%%%%%%%%%%%%%%
\begin{figure}[t]
    \centering
    \includegraphics[width=0.99\linewidth]{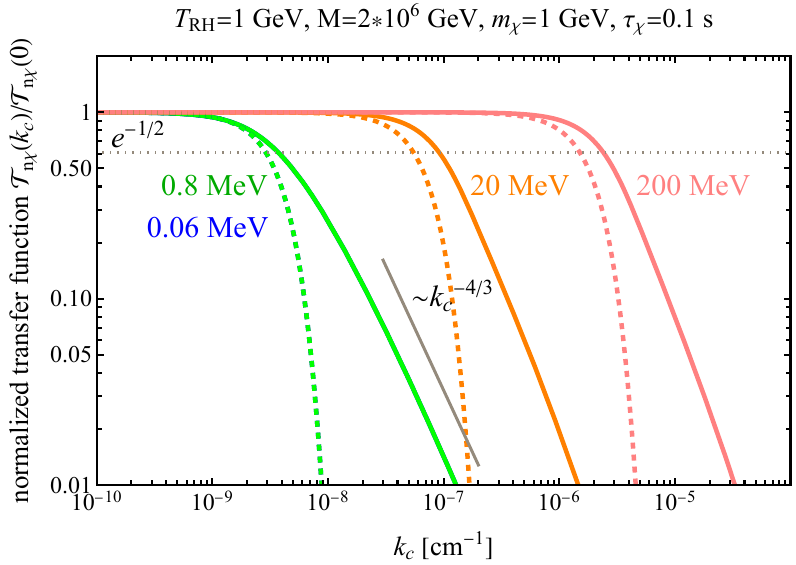}
    \caption{Comparison of the damping at small scales for $\chi$ particles and baryons, at temperatures $T=200\MeV$, $20\MeV$, $0.8\MeV$, and $0.06\MeV$. Solid lines show the transfer function ${\cal T}_{n\chi}(k_c,t,t_{\rm RH})$ normalized by its value at $k_c=0$, and dashed lines the Gaussian damping $\exp(-k_c^2L_c(t)^2)$ of $\chi$. The curves for $0.8\MeV$ and $0.06\MeV$ coincide, since $\chi$ has decayed by then and the subsequent neutron diffusion is negligible on these scales. The horizontal dotted line indicates the threshold $e^{-1/2}$ used in the criterion~\eqref{eq:criteria-cutoff}, and the gray line shows the asymptotic $k_c^{-4/3}$ scaling. Although the solid curves fall off as a power law rather than a Gaussian, both cutoffs lie close to each other at this threshold, so that the analytic estimate of Sec.~\ref{section:diffusiveTreatment} agrees well with the full treatment. We take $T_{\rm RH}=1\,{\rm GeV}$, $M=2\times10^6\,{\rm GeV}$, $m_\chi=1\,{\rm GeV}$, and $\tau_\chi=0.1\,{\rm s}$, and normalize the scale factor to unity at $T=1\,{\rm MeV}$.}
    \label{fig:trf}
\end{figure}
%%%%%%%%%%%%%%%%%%%%%%%%

Let us now explain the qualitative behavior of the transfer function. In Fig.~\ref{fig:trf}, we plot the Gaussian damping factor of $\chi$, $\exp(-k_c^2 L_c(t)^2)$ in Eq.~\eqref{eq:nchi}, with $L_c(t)=L_{\rm erase}(t)/a(t)$ obtained from Eq.~\eqref{Eq:Lerase} (dashed lines), together with the damping factor of neutrons, ${\cal T}_{n\chi}(k_c,t,t_{\rm RH})/{\cal T}_{n\chi}(0,t,t_{\rm RH})$, obtained by numerically integrating Eqs.~\eqref{eq:def-trf} and~\eqref{eq:ltilde} (solid lines). Comparing these two suppression factors, one may recognize the following features. At large scales, $k_c\ll L_c(t)^{-1}$, $\chi$ diffusion is too slow to smear out the initial inhomogeneity, implying no $k_c$-dependent suppression. On the other hand, at smaller scales, $k_c \gtrsim L_c(t)^{-1}$, $\chi$ diffusion matters, which is common to the damping factors of both $\chi$ particles and neutrons. For the inhomogeneity in the dark sector, this results in the Gaussian damping $\exp(-k_c^2 L_c(t)^2)$.

By contrast, the baryon inhomogeneity exhibits a power-law damping, which is due to the inhomogeneities of the neutrons produced by earlier $\chi$ decays. To understand this feature more quantitatively, we examine the behavior of the transfer function. Approximating $\tilde X_n(t')\simeq1/2$ and $k_c^2\tilde L_c(t,t')^2\simeq k_c^2L_c(t')^2\gg(t'-t_{\rm RH})/\tau_\chi$, we obtain
    \begin{align}
        {\cal T}_{n\chi}
        &\simeq\dfrac{1}{2\tau_\chi}\int_{t_{\rm RH}}^t dt'\exp(-k_c^2 L_c(t')^2)\notag\\
        &=\dfrac{1}{2\tau_\chi}\int_{0}^{u(t)}du \dfrac{dt'}{du}e^{-u^2} \, ,
        \label{eq:Tnchi-ana}
    \end{align}
where $u(t)\equiv k_cL_c(t)$. Although $k_c$ enters both $dt'/du$ and the upper limit $u(t)$, the factor $e^{-u^2}$ renders the integral insensitive to the latter. Therefore, for a power law $L_c(t) \propto t^n$ with a general index $n$, we obtain ${\cal T}_{n\chi}\sim k_c^{-1/n}$. In the diffusive regime, Eqs.~\eqref{Eq:LeraseAna} and~\eqref{Eq:timeTemp} give $n=3/4$ up to the variation of $g_*$, so that Eq.~\eqref{eq:Tnchi-ana} implies ${\cal T}_{n\chi}\sim k_c^{-4/3}$. In the free-streaming regime, on the other hand, the comoving length $L_c(t)=d_{\rm fs}(t)/a(t)$ grows only logarithmically, $L_c(t)\propto\ln(t/t_{\rm RH})$, up to the variation of $g_*$, as follows from Eq.~\eqref{Eq:FreeStreamingLength}. This leads to $dt'/du\propto k_c^{-1}$ up to minor corrections at large $k_c$, and hence ${\cal T}_{n\chi}\sim k_c^{-1}$.\footnote{The same power-law suppression can arise even when $N_{\rm tot}<\max[m_\chi/T_{\rm RH},1]$. As the simplest example, consider a monochromatic velocity distribution $f_0(p_c)= \delta(p_c -p_0)/(4\pi p_0^2)$ with $p_0\neq 0$. In this case, the $\chi$ distribution is not damped by a Gaussian, but by the factor $j_0(k_c\ell_{\rm fs}(p_0,t))$ in Eq.~\eqref{Eq:BesselSol}, so that $e^{-u^2}$ in Eq.~\eqref{eq:Tnchi-ana} is replaced by $j_0(u)$. Even then, we find ${\cal T}_{n\chi}\sim k_c^{-1}$ on scales below $\ell_{\rm fs}(p_0,t)$. The situation may change, however, when low-momentum modes dominate, i.e.\ when $4\pi p_c^2 f_0(p_c)$ is peaked at $p_c\simeq 0$. In the extreme case $p_0=0$ of the above example, there is no free-streaming suppression at all, and only the second term in Eq.~\eqref{eq:ltilde} remains. The discussion of Refs.~\cite{Applegate:1987hm,Bagherian:2025puf} then applies as it is, without any benefit from dark transport.}
The damping of the baryon inhomogeneity at small scales is thus a power law, much less efficient than the Gaussian damping in the dark sector, as illustrated by the gray guide line in Fig.~\ref{fig:trf}. Physically, this is because the baryon distribution accumulates the history of the small-scale inhomogeneities transferred from the dark sector at earlier times. Note that these indices apply only for $T\ll{\cal O}(100)\,{\rm MeV}$, where $g_*(T)$ stays approximately constant.

\begin{figure}[t]
    \centering
    \includegraphics[width=0.99\linewidth]{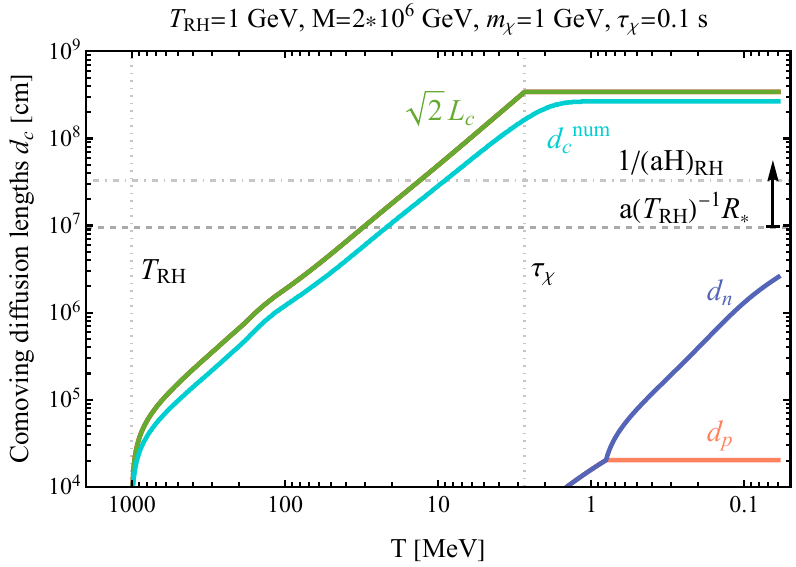}
\caption{Comparison of comoving damping length scales as a function of the temperature $T$. The analytic estimate of the damping scale, $\sqrt{2} L_c(T) = \sqrt{2}L_{\rm erase}(T)/a(T)$ from Eq.~\eqref{Eq:Lerase}, is shown by the green solid line, and the numerically evaluated damping scale $d_c^{\rm num}(T)$, obtained from the transfer function ${\cal T}_{n\chi}(k_c,t,t_{\rm RH})$ in Eq.~\eqref{eq:def-trf}, by the cyan solid line. These agree well with each other. We take $M=2\times10^6 \GeV$, $m_\chi=1 \GeV$, and $\tau_\chi=0.1~\mathrm{s}$, which lie in the diffusive regime of Fig.~\ref{fig:Mbnd}, and normalize the scale factor to unity at $T=1 \MeV$. The gray dot-dashed and dashed lines show the comoving Hubble length at reheating, $(aH)_{\rm RH}^{-1}$, and the initial inhomogeneity scale, $R_*/a(t_{\rm RH})$, respectively, with $T_{\rm RH}=1\GeV$. To erase the inhomogeneity, the baryon damping scale must lie above the gray dashed line at the time of deuterium formation, as indicated by the black arrow. For our parameter choice, this is achieved with the help of $\chi$ diffusion. For comparison, the standard neutron and proton diffusion lengths, $d_n(T)$ and $d_p(T)$, are shown by the blue and red lines, respectively, assuming instantaneous weak decoupling~\cite{Applegate:1987hm,Bagherian:2025puf}. Evidently, these are insufficient to erase the inhomogeneities created by the low-scale PT.}
\label{fig:scales}
\end{figure}

We now define the comoving damping scale of baryons, $d_c^{\rm num}(t)$, as the numerically determined scale such that
\begin{align}
    \dfrac{{\cal T}_{n\chi}(d_c^{\rm num}(t)^{-1},t,t_{\rm RH})}{{\cal T}_{n\chi}(0,t,t_{\rm RH})}=e^{-1/2}\,,
    \label{eq:criteria-cutoff}
\end{align}
slightly modifying the criterion of Ref.~\cite{Bagherian:2025puf}. Fig.~\ref{fig:scales} shows the temperature evolution of the analytic damping scale of $\chi$, $\sqrt{2} L_c(t)$ (green solid line), and of the baryon damping scale $d_c^{\rm num}(t)$ (cyan solid line), for the benchmark parameters $M=2 \times 10^{6} \GeV$, $m_\chi = 1 \GeV$, and $\tau_\chi=0.1 \, {\rm s}$, which lie in the diffusive regime of Fig.~\ref{fig:Mbnd}. The analytic estimate agrees with $d_c^{\rm num}(T)$ within an $\mathcal{O}(1)$ factor at all temperatures, which supports using the analytically estimated $\chi$ erasure length of the previous sections as the damping scale of the baryon distribution. The numerical scale $d_c^{\rm num}(T)$ is slightly smaller than $\sqrt{2} L_c(T)$, owing to the small-scale structure accumulated from $\chi$ particles that decayed at earlier times, while neutron diffusion after the $\chi$ decay remains negligible compared with $d_c^{\rm num}(T)$.

We note, however, that since the cutoff of the baryon distribution is a power law rather than a Gaussian, the two scales would gradually separate if a lower threshold were adopted. This is relevant when the initial inhomogeneity of $\chi$ is large, e.g.\ $\delta n_\chi/\bar n_\chi \sim \mathcal{O}(10^2)$.

%%%%%%%%%%%%%%%%%%%%%%%%%%%%%%%%%%%
%%%%%%%%%%%%%%%%%%%%%%%%%%%%%%%%%%%

%%%%%%%%%%%%%%%%%%%%%%%%%%%%%%%%%%%%%%%%%%%%%%%%%%%%%%%
\subsection{An example model for $\chi$ interaction}
%%%%%%%%%%%%%%%%%%%%%%%%%%%%%%%%%%%%%%%%%%%%%%%%%%%%%%%

In this subsection, we consider a specific model for $\chi$-scattering and show how the scale $M$ is related to the parameters of this model. Note that Eq.~\eqref{Eq:lmfp} defines the effective scale $M$ in terms of the total elastic scattering cross section for $\chi$, denoted as $\langle \sigma v_\mathrm{rel}\rangle_\chi$. If Eq.~\eqref{Eq:portal} is the only interaction that $\chi$ possesses, then the effective induced cross section for elastic scattering with SM neutrons can be estimated as~\cite{Girmohanta:2026eqg} %follows
\begin{align}
\langle \sigma v_\mathrm{rel}\rangle_{\chi n \to \chi n} \simeq \frac{\Lambda_{\rm QCD}^{10}}{\Lambda_n^8 m_\chi^4} \, ,
\label{Eq:chinScattering}
\end{align}
where $\Lambda_{\rm QCD} \simeq 0.2\,{\rm GeV}$ denotes the QCD scale. Furthermore, as the number density of neutrons is suppressed relative to the number density of photons by a factor of $\eta$, the effective scale $M$ comes out to be $\sim 10^{28}\,{\rm GeV}$ for $\tau_\chi \simeq 0.1\,{\rm s}$ and $m_\chi \simeq 1\,{\rm GeV}$. This lies in the free-streaming regime and may satisfy the bound presented in Fig.~\ref{fig:Mbnd}, provided that, $\chi$ acquires a sufficient velocity dispersion by some other means after reheating, e.g., via its interaction with the bubble wall. 
Possible scatterings of $\chi$ with DM are also suppressed due to the small number density of DM compared to radiation. 
Therefore, an additional interaction between $\chi$ and light particles in the thermal bath is desirable, which must nevertheless be weak enough for the erasure length of the baryon structure to be sufficiently long. %to obtain the most stringent bound, 
We will consider an example model in which $\chi$ scatters with a light DS scalar $\varphi$ in the thermal bath through a Yukawa-type interaction, and translate the constraints obtained in the previous subsections into constraints on the Yukawa coupling.

The relevant interaction is described by the following Lagrangian
\begin{align}
-{\cal L}_{\rm int} \supset \mathcal{Y}_{\chi}\,\overline{\chi}\,\chi\,\varphi \, .
\label{Eq:chiInt}
\end{align}
We consider the scattering process $\chi\varphi\to\chi\varphi$ mediated by $\chi$. Since diffusion is dominated at late times, the effective cross section is approximately temperature independent and scales as
\begin{align}
\nonumber
\langle \sigma v_\mathrm{rel} \rangle_\chi &= \langle \sigma v_\mathrm{rel}\rangle_{\chi n \to \chi n}+ \langle \sigma v_\mathrm{rel}\rangle_{\chi \varphi \to \chi \varphi} \\
                              &= \frac{\pi^2}{g_\varphi \zeta(3)} \frac{1}{M^2}
\simeq \frac{|\mathcal{Y}_{\chi}|^4}{8\pi m_\chi^2} \, ,
\label{Eq:sigmaApprox}
\end{align}
where the neutron portal induced contribution in Eq.~\eqref{Eq:chinScattering} is negligible. The erasure condition in Eq.~\eqref{Eq:Mbound} is then satisfied for
\begin{align}
|\mathcal{Y}_{\chi}| \lesssim 10^{-2}
\left(\frac{m_\chi}{1\,{\rm GeV}}\right)^{1/2} \, ,
\end{align}
where we have taken $\tau_\chi=0.1\,{\rm s}$, $T_{\rm RH}=1\,{\rm GeV}$, $g_\varphi=1$, and $\beta/H_{\rm RH}=10$. Furthermore, for $|\mathcal{Y}_{\chi}|\gtrsim 10^{-4}$, the interaction in Eq.~\eqref{Eq:chiInt} alone ensures $N_{\rm tot}\gtrsim\max[m_\chi/T_{\rm RH},1]$ with the same benchmark parameters, which is sufficient for $\chi$ to acquire an MB distribution right after reheating. Meanwhile, the decay of $\varphi$ into SM particles before the onset of BBN can be arranged through the introduction of a Higgs portal.

Finally, we comment on the $n\leftrightarrow\chi$ conversion mediated by the neutron portal, which could keep the two species in chemical equilibrium and require a coupled treatment, in analogy with the $n\leftrightarrow p$ system before weak decoupling. Evaluating the relevant $2\to2$ rates, with $\varphi$ scattering below the QCD scale and quark scattering above it, we find that such equilibrium is established below the reheating temperature if $\tau_\chi$ lies in the interval $10^{-6}\,{\rm s}\lesssim\tau_\chi\lesssim10^{-5}\,{\rm s}$, where the quark channel is active, and we have taken the benchmark $m_\chi \sim T_{\rm RH} \sim \GeV$. For longer lifetimes the conversion proceeds through $\varphi$ and is suppressed by $|\mathcal{Y}_\chi|^2$, so that $\chi$ has already chemically decoupled and retains the entire asymmetry, as assumed above. Even within that window our conclusions are unaffected, and in fact strengthened, because, just as the transport of the coupled $n$--$p$ fluid is dominated by neutron diffusion, that of the coupled $\chi$--$n$ system is dominated by $\chi$, so that the erasure remains Gaussian rather than power law, until decoupling.

%%%%%%%%%%%%%%%%%%%%%%%%%%%%%%%%%%%%%%%%%%%%%%%%%%%
\section{Conclusions}\label{conclusions}
%%%%%%%%%%%%%%%%%%%%%%%%%%%%%%%%%%%%%%%%%%%%%%%%%%%

We have shown that DS transport provides a natural resolution to the baryon-inhomogeneity problem afflicting low-scale supercooled PTs. We have considered a darkogenesis setup, where the asymmetry resides in an SM-singlet Dirac fermion $\chi$ for a period $\tau_\chi$ before being transferred to visible baryons. Thus, the evolution of the inhomogeneity imprinted at the bubble-collision scale $R_*$ is governed by $\chi$ transport, instead of the much slower SM baryon diffusion. 
Introducing an interaction of $\chi$ with the thermal plasma and 
parameterizing its total scattering cross section by a single scale $M$, defined through Eq.~\eqref{Eq:lmfp}, we analyzed both the diffusive and collisionless transport regimes using the Boltzmann equation for the $\chi$ phase-space distribution. In the diffusive case, the inhomogeneity erasure is controlled by the diffusion length $L_{\rm erase}$ in Eq.~\eqref{Eq:LeraseAna}, yielding a lower bound on the scale $M$. In the collisionless case, the erasure is instead governed by the velocity dispersion of $\chi$, which drives a similar exponential suppression, and the corresponding condition is readily satisfied for $\tau_\chi\gg t_{\rm RH}$. A viable parameter region therefore exists in which the inhomogeneity is erased, as illustrated by the white region in Fig.~\ref{fig:Mbnd}.

We note, however, that our main results in Fig.~\ref{fig:Mbnd} are based on the analytic estimate in Eq.~\eqref{Eq:eraseCond}, together with Eq.~\eqref{Eq:LeraseAna}, which is motivated by the exponential damping of the inhomogeneities in the $\chi$ distribution. As shown in Sec.~\ref{sec:baryondamp}, where we go beyond the sudden-decay approximation, the baryon inhomogeneities are instead suppressed as a power law. This difference does not significantly alter our results as long as the initial inhomogeneity in the $\chi$ distribution is at most of order unity. For models with much larger initial inhomogeneities, on the other hand, the Boltzmann equations need to be solved carefully in a model-dependent manner. We also note that, in the free-streaming regime, even when a thermal velocity dispersion does not develop after reheating, DS transport can still suppress the baryon inhomogeneity as a power law, unless the $\chi$ momentum distribution is peaked at zero momentum. Such suppression may still be sufficient to erase the unwanted inhomogeneities and ensure successful BBN, depending on the detailed structure of the initial distribution function. Detailed studies of these model-dependent issues are beyond the scope of this work, but may become important when applying our analysis to specific models of supercooled PTs.

Finally, we emphasize that a low-scale, strongly supercooled PT that does not produce the baryon asymmetry and DM after the transition, but instead relies on a large pre-existing asymmetry diluted to the observed abundance, may acquire additional baryon inhomogeneities from the inhomogeneous reheating by the bubble dynamics, on top of the homogeneous pre-existing abundance. These inhomogeneities are difficult to erase, since baryon diffusion is slow for reheating temperatures below the TeV scale. In contrast, in a strongly supercooled PT-assisted darkogenesis setup, where the baryon asymmetry and DM are generated only after the pre-existing asymmetry has been reduced to negligible values through entropy dilution, the resulting inhomogeneities can be efficiently erased through dark-sector transport. In light of the precise measurements of the deuterium abundance D/H, this makes darkogenesis a particularly well-motivated scenario, especially for low-scale supercooled PTs invoked to explain the PTA signal.
\\\\

\section*{Acknowledgments}
SG thanks Tae Hyun Jung for useful discussion. SG acknowledges support by IBS under the project code IBS-R018-D1.
The work of KK was supported
by the National Natural Science Foundation of China
(NSFC) under Grant Nos.~W2532007 and 12547104, and
by JSPS KAKENHI Grant-in-Aid for Challenging Research (Exploratory) JP23K17687.
YN is supported by Natural Science Foundation of Shanghai.
FU is supported by IBS under the project code IBS-R018-D3.

\bibliographystyle{JHEP}
\bibliography{ref}

\end{document}